\documentclass[a4paper,12pt]{article}
\usepackage[utf8]{inputenc}
\usepackage[english]{babel}
\usepackage{amsmath, amssymb, graphicx,bm}
\usepackage{braket}
\usepackage{caption}
\usepackage{subcaption}
\usepackage{ulem}
\usepackage{multicol}

\pdfoutput=1
\usepackage{jheppubm}
\newcommand{\be}{\begin{equation}}
\newcommand{\ee}{\end{equation}}
\newcommand{\bea}{\begin{eqnarray}}
\newcommand{\eea}{\end{eqnarray}}

\numberwithin{equation}{section}

\begin{document}

\title{
Information Problem in Black Holes and Cosmology and Ghosts in Quadratic Gravity}

\author{Igor Volovich}

\affiliation{Steklov Mathematical Institute, Russian Academy of
  Sciences,\\ Gubkina str. 8, 119991, Moscow, Russia}

\emailAdd{volovich@mi-ras.ru}

\date{January 2024}
\abstract{
Black hole information problem is the question about unitarity of the evolution operator   during the collapse and evaporation of the black hole.  One can ask the same question about unitarity of quantum and inflationary cosmology. In this paper we argue that in both cases, for black holes and for cosmology, the answer is negative and we face non-unitarity.

  Such a question  can not be addressed by using the fixed classical gravitational background since one has to take into account the backreaction. To his end one  uses the semi-classical gravity, which includes the expectation value of the energy - momentum tensor operator  of the matter fields. One has to renormalize the energy-momentum tensor and one gets an effective action which contains quadratic terms in  scalar curvature and  Ricci tensor. Such quadratic gravity contains ghosts which in fact lead to violation of unitarity in black holes and cosmology. We discuss the question whether black holes will emit ghosts. 
  
  One can try to restrict ourselves to the  $f(R)$ gravity that seems is a good approximation to the semi-classical gravity  and widely used in cosmology.  The black hole entropy in $f(R)$ gravity is different from the Bekenstein-Hawking entropy and  from entanglement island entropy. The black hole entropy in $R+R^2$ gravity  goes to a constant during the evaporation process. This can be interpreted as  another indication to the possible non-unitarity in black holes and cosmology.  
}


\maketitle



\section{Introduction}

Black hole information problem is the question about unitarity of the evolution operator  during the collapse and evaporation of the black hole \cite{Hawking:1976ra,FroNov}.  Various proposals have been discussed in this context, such as black hole explosion \cite{Hawking:1974rv,Arefeva:2021byb},  gravitational collapse of quantum scalar field \cite{Berczi:2020nqy}, complementarity \cite{Susskind:1993if}, firewalls \cite{Almheiri:2012rt}, islands \cite{Penington:2019npb, Almheiri:2019psf,Almheiri:2019yqk, Hashimoto:2020cas, Arefeva:2021kfx, Balasubramanian:2021xcm, 
Stepanenko:2022gwy}, near critical state equations \cite{Arefeva:2022guf, Arefeva:2022cam} and negative dimensions \cite{Arefeva:2023kpu, Arefeva:2023kwm}.

If one considers a fixed classical gravitational background then we cannot justify discussion of the black hole evaporation and therefore there is no reason to speak about information problem in such approximation. One can try to attack information loss problem if we would investigate the collapse of quantum field or discuss semi-classical gravity equations. These equations include  the expectation value of the energy-momentum tensor operator  of the matter fields. One has to renormalize the energy-momentum tensor and one gets an effective action which contains higher order terms in curvature including not only linear term in curvature, but also quadratic in scalar and Ricci tensor. Such quadratic gravity is renormalizable but it contains ghosts \cite{Stelle:1976gc,Salvio:2018crh}. These ghosts lead to violation of unitarity in black holes and cosmology. We discuss the question whether black holes  will emit ghosts.

In this note, in the context  of the black hole information problem,   we consider also an  effective classical local $f(R)$ gravity which seems is a  rather good approximation to the semi-classical gravity.     In a simple case it describes an interaction of gravity with quantum conformal matter.
$f(R)$ gravity is wildly used in inflationary cosmology \cite{Starobinsky:1980te, Vilenkin:1985md, Linde:1990xn}. Black holes in $f(r)$ gravity are considered in \cite{Nojiri:2010wj}.

\subsection{Classical and semi-classical gravity}
Hawking radiation has been derived by considering a free quantum field in the classical gravitational background.
For example, the Schwarzschild black hole metric satisfies the Einstein equation 
\be
R_{\mu\nu}-\frac{1}{2}R g_{\mu\nu}=0\ee
In such an approach we cannot investigate the black hole information problem since during the evaporation process we have to taken into account the back  reaction of the radiation to the metric. To this end usually one  uses the semi-classical gravity approximation.

 Semi-classical gravity is the approximation to the theory of quantum gravity in which one treats matter fields as being quantum and the gravitational field as being classical but including quantum corrections.

  The curvature of the spacetime is given by the semi-classical Einstein equations, which relate the curvature of the spacetime to the expectation value of the energy-momentum tensor operator $T_{\mu \nu }$ of the matter fields:
\begin{equation}
\label{expectvalue}
R_{\mu\nu}-\frac{1}{2}R g_{\mu\nu} =  8 \pi G \left\langle \hat T_{\mu\nu} \right\rangle_\psi \,,
 \end{equation}
 where G is the gravitational constant and 
$\psi$  indicates the quantum state of the matter fields.
 Such equations are non-linear  and non-local equations that are difficult to analyze. Therefore  we consider the effective gravity.

In cosmology with the Friedmann metric
\be
ds^2=-dt^2+a(t)^2 d\Omega
\ee
the semi-classical gravity with conformal matter was considered on \cite{Starobinsky:1980te} and is called the Starobinsky inflationary model. It is shown in \cite{ Vilenkin:1985md} that this model in 1-loop approximation is equivalent to the effective $R+R^2$ gravity.

By analogy with cosmology we will consider the $R+R^2$ gravity as the approximation to the semi-classical gravity in theory of black holes and use it to describe the evaporation process.

The black hole entropy in $f(R)$ gravity is different from the Bekenstein-Hawking entropy and from the island entropy. It  goes to a constant during the evaporation process. This can be interpreted as an indication to the possible none-unitary evolution of the system. 

\section{Semi-classical gravity and quadratic gravity }
For the semi-classical Einstein equation \eqref{expectvalue} we have to perform  renormalizations in both parts. 
The renormalized  version of the left hand site of  Eq.\eqref{expectvalue} reads  (see Eq.(6.52) \cite{Birrell:1982ix})
\begin{equation}
    R_{\mu\nu} - \frac{1}{2}Rg_{\mu\nu} + \alpha ^{(1)}H_{\mu\nu} + \beta ^{(2)}H_{\mu\nu} + \gamma H_{\mu\nu},
\end{equation}
where
\begin{equation}
    ^{(1)}H_{\mu\nu}= 2R_{;\mu\nu}-2g_{\mu\nu}\Box R - \frac{1}{2}g_{\mu\nu} R^2 + 2R R_{\mu\nu},
\end{equation}
\begin{equation}
     ^{(2)}H_{\mu\nu}= 2R_{\mu;\nu \alpha} ^{\alpha} -\Box R_{\mu\nu} - \frac{1}{2}g_{\mu\nu}\Box R + 2R_{\mu}^{\alpha}R_{\alpha\nu}- \frac{1}{2} g_{\mu\nu} R^{\alpha\beta}R_{\alpha\beta}
\end{equation}
\begin{equation}
    H_{\mu\nu}= -\frac{1}{2} g_{\mu\nu} R^{\alpha\beta \gamma \delta}R_{\alpha\beta\gamma \delta} + 2R_{\mu\alpha\beta \gamma} R_{\nu}^{\alpha\beta \gamma} 
  -4   \Box R_{\mu\nu} 
    + 2R_{; \mu \nu }- 4R{\mu \alpha} R^{\alpha} _{\nu} + 4R^{\alpha \beta} R_{\alpha \mu \beta \nu},
\end{equation}
where $\alpha, \beta, \gamma$ are constant. 

One has the relations
\be
 ^{(1)}H_{\mu\nu}=-\frac{1}{\sqrt{-g}}\frac{\delta }{\delta g^{\mu \nu}(x)}\int dx \sqrt{-g}R^2
   \ee
   and similar for the square of the scalar curvature and the Ricci tensor.  It is known that the divergences in the energy-momentum tensor can be remover by usin the regularization of the type $\left\langle\phi(x)\phi(y)\right\rangle$ as $x\to y$. 
So one get the renormalized action in the form
\begin{equation}
    S_{ren}= \int d^4x \sqrt{-g} (16\pi G)^{-1} (R-2\Lambda) + aR^2 + bR^{\alpha \beta}R_{\alpha \beta} + cR^{\alpha \beta \gamma \delta} R_{\alpha \beta \gamma \delta}
\end{equation}
This action describes quadratic gravity. 
It is known quadratic gravity is renormalizable field theory, but it contains ghosts \cite{Stelle:1976gc,Donoghue:2021cza}. That this ghost lead to the violation of the unitarity theory. Therefore the semi-classical also should lead to the violation of unitarity. 

A natural question arises whether the black hole will emit ghost by mechanism similar to the usual Hawking radiation. Lets remand that in classical field theory with higher derivative there are not with negative energy. Which after the quantization can be represented as $S_{ghost}$. In both cases, classical and quantum theories, this means a lack of stability. Whether we have ghostly radiation from a black hole is a question worthy of further investigation. 

\section{Effective $f(R)$ gravity.}
Let us consider the gravitational field with matter in the following form $ I=I_{EH}+ I_{mat}$
\begin{equation}
    I_{EH}=\frac{1}{16\pi G} \int d^4 x \sqrt{-g} R,
\end{equation}
where $g_{\mu\nu}$ is a metric, $g=\det g_{\mu\nu} $ and $R$ is a Ricci scalar. 
If we consider a scalar field, then
the action for the matter field $I_{mat}$ has the following form
\begin{equation}
    I_{mat}=\frac{1}{2} \int d^4 x \sqrt{-g} (\xi \varphi^2 R-g^{\mu\nu}\partial_\mu \varphi \partial_\nu \varphi ),
\end{equation}
where $\xi$ is a constant.

Partition function in quantum gravity is given by the Euclidean path integral
\begin{equation}
Z=\int Dg_{\mu\nu} D\varphi \,e^{-I }=\int  D g_{\mu\nu} \,e^{-I_{eff}(g)}
\end{equation}
One can expand $I_{eff}(g)$ as follows
\begin{equation}
    I_{eff}(g) = f(R) + f_1(R_{\mu\nu}R^{\mu\nu}) + f_2(R_{\mu\nu\alpha\beta}R^{\mu\nu\alpha\beta})+ f_3 (\Box R) + ...
\end{equation}
The action of  $f(R)$ gravity is
\cite{Ferraris:1993de,Capozziello:2002rd,Capozziello:2003tk,Carroll2004, Faraoni:2010yi}
\be\label{action1}
I=\frac{1}{16\pi G} \int d^4 x \, \sqrt{-g} \, 
f(R),
\end{equation}
where $f(R)$ is analytical or smooth function. Variation of the action with respect to the 
metric  $g^{\mu\nu}$ gives the field equations
\be\label{metriv-eom}
f'(R)R_{\mu\nu}-\frac{f(R)}{2} \, 
g_{\mu\nu}=\nabla_{\mu}\nabla_{\nu} f'(R) -g_{\mu\nu} \Box 
f'(R),
\end{equation}
where a prime denotes differentiation with respect to $R$.  

Vacuum equations of motions are

\be
 R_{\mu\nu}(f'(R))-\frac{1}{2}\,g_{\mu\nu}\,(f(R))\,=\,0
\ee
The 
constant scalar curvature solutions $R\,=\,R_0$ means:
\begin{eqnarray}
 2f'(R_0)\,R_{0}-4\,f(R_{0})\,=\,0
\label{root_R0}
\end{eqnarray}
For this kind of solution an effective cosmological constant may be defined
 as $\Lambda_D^{eff}\equiv R_{0}/D$,  for a recent derivation of the inflationary cosmological constant, see
  \cite{Volovich:2023vib}. Thus, for any solution with constant curvature
$R=R_0$ and $f'(R_0)\neq 0$ holds:

\begin{eqnarray}
 R_{\mu\nu}=\frac{f(R_0)}{2f'(R_0)}\,g_{\mu\nu}
\end{eqnarray}

\section{Black Holes in $f(R)$ gravity}
 The general static spherically symmetric solution can be written as
\be
ds^2\,=- h (r) dt^2+h^{-1}(r)d
r^2+r^2d\Omega^2,
\label{metric_D_v1}
\ee
where
\be
h(r)\,=\,1-\frac{2 G M}{r}+\frac{Q^2}{r^{2}}.
\label{RN}
\ee
This solution corresponds to a Reissner-Nordstr\"{o}m solution, i.e. a charged massive BH solution with mass $M$
and charge $Q$. Below we restrict ourselves to the Schwarzschild solution, i.e. $Q = 0$.

Problem of ghosts in $R^2$ gravity is analyzed in \cite{Alvarez-Gaume:2015rwa, Hell:2023mph}.
Note  that there are ghosts in $R+R^2$ gravity in the Schwarzschild background. Consider perturbations around the Schwarzschild background where $g_{\mu\nu} = \overline{g}_{\mu\nu}+ h_{\mu\nu}$ is the Schwarzschild solution with $\overline{R}=0$. Then the $R^2$ action one can consider in the following form
\begin{equation}
    S=\int d^4 x \sqrt{g} R^2.
\end{equation}
Then the $R^2$ action  is
\begin{equation}
    S_2=b\int d^4 x[D_\mu D_\nu h^{\mu\nu}- \Box h ]^2,
\end{equation}
where $D_\mu$ is covariant derivative with respect to the Schwarzschild metric.
We see the appearance of ghost in this action, similarly to the appearance of ghosts in Minkowski background. If we add to the pure $R^2$ action other linear scalar term then we get propagating ghosts.
Note that in the de-Sitter black hole there are no ghosts, see \cite{Sbisa:2014pzo}.

\section{The entropy in $f(R)$ gravity}

The entropy of the black hole in the $f(r)$ gravity  is
\cite{Wang:2005bi, Vollick:2007fh, Faraoni:2010yi, Nojiri:2010wj}
\be\label{metricf(R)entropy}
S=\frac{ f'(R_H)A}{4G},
\ee
where $A=16 \pi G^2M^2$ is the area of the horizon of the black hole  with  mass $M$  and $R_H$ is the curvature on the horizon. For the Schwarzschild black hole one has
\be
R_H = \frac{2}{r_h^2}=\frac{1}{2G^2 M^2}.
\ee
This formula for entropy was obtained by the Noether charge method proposed by Wald
\cite{Wald:1993nt}
\be
S = 4\pi \int_{S^2} \frac{\partial {\cal L}}{\partial R} d^2 x,
\ee
where ${\cal L}$ is the Lagrangian density and the integral is over the horizon at $r=r_h$.
We consider
\be
f(R)=R+\lambda G  R^2,
\ee
where $\lambda$ is dimensionless number. Finally, for the entropy of the black hole we have
\begin{equation}
S =\frac{ f'(R_H)A}{4G} = \frac{(1+\frac{\lambda}{G M^2})16 \pi G^2M^2}{4G}=4\pi G M^2 + 4\lambda.
\end{equation}
So, the entropy for effective $R+R^2$ gravity is equal to the sum of the Bekenstein-Hawking entropy
\begin{equation}
    S_{BH}=\frac{A}{4G}= 4 \pi G M^2
\end{equation}
and the constant $4\lambda$. If we apply this formula to the evaporation process, when the mass of the black hole $M$ decreases, after the complete evaporation we have the non-vanishing entropy $S = 4 \lambda$.

\section{Conclusions}
It is pointed out in this paper that the semi-classical Einstein equation after renormalization are reduced to the quadratic gravity that contains ghosts. The presence of ghosts violates unitarity of theory, in other words there is information lost in black holes in cosmology. In that situation arises question whether black holes will emit ghosts by mechanism similar to the usual Hawking radiation. 

The effective $f(R)$ gravity is also considered in the context of the black hole information problem. 
It is noted that, since this problem must be studied taking into account collapse and evaporation processes, one cannot limit oneself to the classical fixed background of a black hole, but it is necessary to take into account the effect of the back reaction of radiation on the metric. 
It is argued that, by analogy with inflationary cosmology, effective $f(R)$ gravity can be used to take into account such a back reaction non-perturbatively.  One considers the Friedmann equation in semi-classical gravity when the expectation value Eq.\eqref{expectvalue} is computed by one-loop approximation. It is shown \cite{Vilenkin:1985md} that in this approximation, quasi-classical gravity can be reduced to $R+R^2$ gravity, which in cosmology is called the Starobinsky inflationary model.

In this paper the conformal invariant gravity $f(R) = R+R^2$ has been considered and it is shown that at the end of  evaporation the black hole still has non-zero constant entropy. 
This raises the question of what complete evaporation of a black hole means even for the Schwarzschild metric.
In the Schwarzschild coordinates one can send the mass of the black hole to zero to get the Minkowski metric, but in the Kruskal coordinates one has the singularity in this limit, see \cite{Arefeva:2021byb, Arefeva:2022guf}. 
It is important to study time-dependent spherically symmetric solutions of $f(R)$ gravity.

In inflationary cosmology the justification of the effective $R+R^2$ gravity is based on experimental results. 
We hope that applying this model to the black hole information problem will also be quite reasonable.

Note that actually the information problem of black holes in cosmology is particular case of fundamental problem of irreversibility and entropy increasing is considered in foundation of statistical mechanics and theory of open quantum systems, see \cite{Nieuwenhuizen:2005zq, Volovich:2009fwa, VolOhya}.

\section{Acknowledgements}
I am very grateful to I.Aref'eva, V.Frolov, T.Rusalev and D.Stepanenko for fruitful discussions.


\begin{thebibliography}{10}

\bibitem{Hawking:1976ra}
S.~W.~Hawking,
 Breakdown of Predictability in Gravitational Collapse,''
Phys. Rev. D \textbf{14} (1976), 2460-2473

\bibitem{FroNov}
V. Frolov, I. Novikov, Black Hole Physics: Basic Concepts and New Developments, Springer, 2012.




\bibitem{Hawking:1974rv}
S.~W.~Hawking,
 Black hole explosions,''
Nature \textbf{248} (1974), 30-31



\bibitem{Arefeva:2021byb}
I.~Aref'eva and I.~Volovich,
 Quantum Explosions of Black Holes and Thermal Coordinates,''
Symmetry \textbf{14} (2022) no.11, 2298,
arXiv:2104.12724 [hep-th].

\bibitem{Berczi:2020nqy}
B.~Berczi, P.~M.~Saffin and S.~Y.~Zhou,
 Gravitational collapse with quantum fields,''
Phys. Rev. D \textbf{104} (2021) no.4, L041703,
arXiv:2010.10142 [gr-qc].

\bibitem{Susskind:1993if}
L.~Susskind, L.~Thorlacius and J.~Uglum,
 The Stretched horizon and black hole complementarity,''
Phys. Rev. D \textbf{48} (1993), 3743-3761,
arXiv:hep-th/9306069 [hep-th].


\bibitem{Almheiri:2012rt}
A.~Almheiri, D.~Marolf, J.~Polchinski and J.~Sully,
 Black Holes: Complementarity or Firewalls?,''
JHEP \textbf{02} (2013), 062,
arXiv:1207.3123 [hep-th].

\bibitem{Penington:2019npb}
G.~Penington,
 Entanglement Wedge Reconstruction and the Information Paradox,''
JHEP \textbf{09} (2020), 002,
arXiv:1905.08255 [hep-th].

\bibitem{Almheiri:2019psf}
A.~Almheiri, N.~Engelhardt, D.~Marolf and H.~Maxfield,
 The entropy of bulk quantum fields and the entanglement wedge of an evaporating black hole,''
JHEP \textbf{12} (2019), 063,
arXiv:1905.08762 [hep-th].


\bibitem{Almheiri:2019yqk}
A.~Almheiri, R.~Mahajan and J.~Maldacena,
 Islands outside the horizon,''
arXiv:1910.11077 [hep-th].


\bibitem{Hashimoto:2020cas}
K.~Hashimoto, N.~Iizuka and Y.~Matsuo,
 Islands in Schwarzschild black holes,''
JHEP \textbf{06} (2020), 085,
arXiv:2004.05863 [hep-th].

\bibitem{Arefeva:2021kfx}
I.~Aref'eva and I.~Volovich,
 A note on islands in Schwarzschild black holes,''
Teor. Mat. Fiz. \textbf{214} (2023) no.3, 500-516,
arXiv:2110.04233 [hep-th].


\bibitem{Balasubramanian:2021xcm} 
V.~Balasubramanian, B.~Craps, M.~Khramtsov and E.~Shaghoulian,
 Submerging islands through thermalization,''
JHEP \textbf{10} (2021), 048,
arXiv:2107.14746 [hep-th].







\bibitem{Stepanenko:2022gwy}
D.~Stepanenko and I.~Volovich,
 Schwarzschild black holes, Islands and Virasoro algebra,''
Eur. Phys. J. Plus \textbf{138} (2023) no.8, 688
arXiv:2211.03153 [hep-th].



\bibitem{Arefeva:2022guf}
I.~Aref'eva and I.~Volovich,
 Complete Evaporation of Black Holes and Page Curves,''
Symmetry \textbf{15} (2023) no.1, 170,
arXiv:2202.00548 [hep-th].

\bibitem{Arefeva:2022cam}
I.~Y.~Aref'eva, T.~A.~Rusalev and I.~V.~Volovich,
 Entanglement entropy of a~near-extremal black hole,''
Teor. Mat. Fiz. \textbf{212} (2022) no.3, 457-477,
arXiv:2202.10259 [hep-th].




\bibitem{Arefeva:2023kpu}
I.~Aref'eva and I.~Volovich,
 Violation of the Third Law of Thermodynamics by Black Holes, Riemann Zeta Function and Bose Gas in Negative Dimensions,''
arXiv:2304.04695 [hep-th].

\bibitem{Arefeva:2023kwm}
I.~Y.~Aref'eva and I.~V.~Volovich,
``Bose Gas Modeling of the Schwarzschild Black Hole Thermodynamics,''
arXiv:2305.19827 [hep-th].
\bibitem{Stelle:1976gc}
K.~S.~Stelle,
``Renormalization of Higher Derivative Quantum Gravity,''
Phys. Rev. D \textbf{16} (1977), 953-969
\bibitem{Salvio:2018crh}
A.~Salvio,
``Quadratic Gravity,''
Front. in Phys. \textbf{6}, 77 (2018),
arXiv:1804.09944 [hep-th]
\bibitem{Starobinsky:1980te}
A.~A.~Starobinsky,
``A New Type of Isotropic Cosmological Models Without Singularity,''
Phys. Lett. B \textbf{91} (1980), 99-102


\bibitem{Vilenkin:1985md}
A.~Vilenkin,
``Classical and Quantum Cosmology of the Starobinsky Inflationary Model,''
Phys. Rev. D \textbf{32} (1985), 2511



\bibitem{Linde:1990xn}
A.~D.~Linde, ``Inflation and quantum cosmology'', AND COSMOLOGY (1990): 537 pp. ed. R.~Brandenberger,




\bibitem{Nojiri:2010wj}
S.~Nojiri and S.~D.~Odintsov,
``Unified cosmic history in modified gravity: from F(R) theory to Lorentz non-invariant models,''
Phys. Rept. \textbf{505} (2011), 59-144,
arXiv:1011.0544 [gr-qc].


\bibitem{Wald:1993nt}
R.~M.~Wald,
``Black hole entropy is the Noether charge,''
Phys. Rev. D \textbf{48} (1993) no.8, R3427-R3431,
[arXiv:gr-qc/9307038 [gr-qc]].

\bibitem{Birrell:1982ix}
N.~D.~Birrell and P.~C.~W.~Davies,
``Quantum Fields in Curved Space,''
Cambridge Univ. Press, 1984,


\bibitem{Donoghue:2021cza}
J.~F.~Donoghue and G.~Menezes,
Nuovo Cim. C \textbf{45} (2022) no.2, 26,
arXiv:2112.01974 [hep-th].

\bibitem{Ferraris:1993de}
M.~Ferraris, M.~Francaviglia and I.~Volovich,
``Universal gravitational equations,''
Nuovo Cim. B \textbf{108}, 1313-1317 (1993)



\bibitem{Capozziello:2002rd}
S.~Capozziello,
``Curvature quintessence,''
Int. J. Mod. Phys. D \textbf{11}, 483-492 (2002),
arXiv:gr-qc/0201033 [gr-qc].

\bibitem{Capozziello:2003tk}
S.~Capozziello, S.~Carloni and A.~Troisi,
``Quintessence without scalar fields,''
Recent Res. Dev. Astron. Astrophys. \textbf{1}, 625 (2003), astro-ph/0303041.



\bibitem{Carroll2004}  Carroll, S.M.,  Duvvuri, V., Trodden, M. and 
Turner, M.S., Is cosmic speed-up due to new gravitational 
physics? {\em Phys. Rev. D}, {\bf 2004} {\em 70}, 043528.

\bibitem{Faraoni:2010yi}
V.~Faraoni, ``Black hole entropy in scalar-tensor and f(R) gravity: An Overview,''
Entropy \textbf{12} (2010), 1246,
arXiv:1005.2327 [gr-qc].



\bibitem{Volovich:2023vib}
I.~Volovich,
``Cosmological Constant and Maximum of Entropy for de Sitter Space,''
arXiv:2308.11377 [hep-th].

\bibitem{Alvarez-Gaume:2015rwa}
L.~Alvarez-Gaume, A.~Kehagias, C.~Kounnas, D.~L\"ust and A.~Riotto,
Fortsch. Phys. \textbf{64} (2016) no.2-3, 176-189
[arXiv:1505.07657 [hep-th]].

\bibitem{Hell:2023mph}
A.~Hell, D.~Lust and G.~Zoupanos,
``On the Degrees of Freedom of $R^2$ Gravity in Flat Spacetime,''
arXiv:2311.08216 [hep-th].

\bibitem{Sbisa:2014pzo}
F.~Sbis\`a,
``Classical and quantum ghosts,''
Eur. J. Phys. \textbf{36} (2015), 015009,
arXiv:1406.4550 [hep-th].

\bibitem{Vollick:2007fh}
D.~N.~Vollick,
``Noether Charge and Black Hole Entropy in Modified Theories of Gravity,''
Phys. Rev. D \textbf{76} (2007), 124001,
arXiv:0710.1859 [gr-qc].


\bibitem{Wang:2005bi}
P.~Wang,
``Horizon entropy in modified gravity,''
Phys. Rev. D \textbf{72}, 024030 (2005),
arXiv:gr-qc/0507034 [gr-qc].



\bibitem{delaCruz-Dombriz:2009pzc}
A.~de la Cruz-Dombriz, A.~Dobado and A.~L.~Maroto,
``Black Holes in f(R) theories,''
Phys. Rev. D \textbf{80}, 124011 (2009)
[erratum: Phys. Rev. D \textbf{83}, 029903 (2011)],
arXiv:0907.3872 [gr-qc].


\bibitem{delaCruz-Dombriz:2012bni}
A.~de la Cruz-Dombriz and D.~Saez-Gomez,
``Black holes, cosmological solutions, future singularities, and their thermodynamical properties in modified gravity theories,''
Entropy \textbf{14}, 1717-1770 (2012)
arXiv:1207.2663 [gr-qc].

\bibitem{Nieuwenhuizen:2005zq}
T.~M.~Nieuwenhuizen and I.~V.~Volovich,
``Role of various entropies in the black hole information loss problem,''
in   Beyond the Quantum. September 2007, 135-145, 
arXiv:hep-th/0507272, 

\bibitem{Volovich:2009fwa}
I.~V.~Volovich,
``Randomness in Classical Mechanics and Quantum Mechanics,''
Found. Phys. \textbf{41} (2011), 516
[arXiv:0910.5391 [quant-ph]].

\bibitem{VolOhya}
Ohya, M., \& Volovich, I. (2011). Mathematical foundations of quantum information and computation and its applications to nano-and bio-systems. Springer Science \& Business Media.

\end{thebibliography}
\end{document}